\documentclass[letterpaper, 10 pt, conference]{ieeeconf}  % Comment this line out if you need a4paper

\IEEEoverridecommandlockouts                              % This command is only needed if 
\usepackage{graphics} 
\usepackage{booktabs}
\usepackage{graphicx}
\usepackage{placeins}

\title{\LARGE \bf
A Tractography Analysis Framework Using Diffusion Maps to Study Thalamic Connectivity in Traumatic Brain Injury }

\author{
    Akul Sharma$^{1}$, Anand A. Joshi$^{1}$, Richard M. Leahy$^{1}$ \\
    \textit{Department of Electrical and Computer Engineering,} \\
    \textit{University of Southern California, Los Angeles, California, USA} \\
    Email: {akulshar@usc.edu}
}

\begin{document}

\maketitle
\thispagestyle{empty}
\pagestyle{empty}

%%%%%%%%%%%%%%%%%%%%%%%%%%%%%%%%%%%%%%%%%%%%%%%%%%%%%%%%%%%%%%%%%%%%
\begin{abstract}

Traumatic brain injury (TBI) disrupts thalamocortical connectivity, contributing to cognitive impairment and post-traumatic epilepsy (PTE). This study presents a novel tractography-based framework that leverages diffusion maps to capture microstructural and organizational changes in thalamic white matter pathways. By analyzing individual streamline characteristics, we identified significant associations between diffusion map embeddings and functional outcomes (GOSE scores), highlighting potential biomarkers for injury severity and recovery trajectories. Our findings suggest that fine-grained geometric features of white matter tracts may provide a more sensitive marker for TBI-related alterations. 
\newline

\indent \textit{Clinical relevance}— This approach enhances the detection of subtle thalamocortical disruptions following TBI, which may improve early identification of patients at risk for poor recovery or PTE. By integrating advanced tractography methods into clinical neuroimaging pipelines, this framework has the potential to refine prognostic models and guide personalized rehabilitation strategies.
\end{abstract}

%%%%%%%%%%%%%%%%%%%%%%%%%%%%%%%%%%%%%%%%%%%%%%%%%%%%%%%%%%%%%%%%%%%%%%%%%%%%%%%%
\section{INTRODUCTION}

Traumatic brain injury (TBI) is an injury to the brain caused by an external force and is one of the leading causes of death and disability in the USA and worldwide. TBI causes cognitive, behavioral, and neurological deficits, and outcomes can range from recovery to permanent disability or death ~\cite{johnson2017}. It can also result in seizures, which is termed as post-traumatic epilepsy (PTE) \cite{jenkinson2012}. Due to its heterogeneous nature, TBI induces complex molecular and neurological changes, posing significant challenges in diagnosis, prognosis, and treatment. TBI causes widespread brain gray and white matter alterations which are associated with poor outcomes ~\cite{vespa2019epibios4rx}.

Diffusion-weighted imaging (DWI) is a powerful tool for investigating white matter architecture in both healthy and pathological conditions \cite{shenton2018}. By measuring the directional diffusion of water molecules within brain tissue, DWI enables the non-invasive mapping of white matter fiber orientations. Diffusion Tensor Imaging (DTI) is the most commonly used model for DWI data, providing quantitative metrics such as Fractional Anisotropy (FA) and Mean Diffusivity (MD), which are widely used to assess white matter microstructure \cite{shenton2018}. However interpreting these metrics, especially in the case of TBI, presents significant limitations as FA reductions can result from demyelination, axonal loss, inflammation, or edema. On the other hand, increased FA may reflect acute injury due to cytotoxic edema rather than improved integrity \cite{garner2019imaging}. In chronic TBI, ongoing neurodegeneration, inflammation, and repair processes further complicate diffusion metric interpretation \cite{xiong2014}. 

Tractography analysis offers a complementary approach that focuses on modeling structural connectivity rather than voxel-wise metrics by reconstructing geometric models of white matter pathways by following fiber orientations estimated from diffusion MRI data \cite{chamberland2019}. With high-quality diffusion imaging, tractography can delineate major white matter bundles, also known as fiber bundles and provide detailed models of brain connectivity. Each reconstructed bundle consists of multiple individual streamlines (fiber tracts), where each streamline is represented as a 3D curve composed of a sequence of points in space. These streamlines approximate the trajectory of axonal pathways connecting different brain regions. While these reconstructions provide geometric information about white matter organization, conventional tractography analyses often reduce these detailed models to simplified bundle-level metrics, potentially overlooking important structural features captured in individual streamline geometry that could be relevant to understanding TBI-related changes \cite{jeurissen2019}. 

However, analyzing tractography data at the streamline level presents a high-dimensional challenge, as each fiber is defined by multiple geometric and positional features that contribute to its spatial organization \cite{jeurissen2019}. Specifically, we characterize each streamline using nine features, including fiber length, curvature, dispersion, and the coordinates of the start, midpoint, and endpoint. These features exhibit strong but nonlinear relationships, making it difficult to directly compare fibers and extract meaningful patterns using conventional dimensionality reduction techniques. Linear methods, such as principal component analysis (PCA), primarily capture variance along orthogonal axes and may fail to optimally represent the feature manifold, particularly in the presence of nonlinear anatomical variations \cite{zhan2015}

To overcome these limitations, we employ diffusion maps, a nonlinear dimensionality reduction technique introduced by Coifman and Lafon \cite{coifman2006}. Diffusion maps construct a sparse graph representation of the data using k-nearest neighbors or an epsilon-ball approach, capturing local geometric relationships between streamlines \cite{coifman2006}. A diffusion operator is then applied to this graph, and its eigenvectors and eigenvalues define a lower-dimensional embedding that transforms the data into an interpretable Euclidean space. A key advantage of diffusion maps is that the Euclidean distances in the embedded space correspond to diffusion distances in the original feature space, ensuring that fibers with similar geometric properties remain close together \cite{coifman2005}. This enables robust clustering of fibers, reduces noise, and preserves intrinsic structural patterns that may be lost in high-dimensional space.

In order to leverage the full geometric complexity of tractography data, we present an analytical framework that analyzes both individual streamline characteristics and intra-bundle connectivity patterns. These geometric and spatial features were used to construct a similarity matrix between fibers, quantifying how similar each fiber is to every other fiber within the bundle. Next, diffusion maps were applied to obtain a lower-dimensional representation that preserves essential structural patterns while reducing redundancy and maintaining anatomical relevance. Using this embedding, we can identify natural groupings of fibers within bundles, detect anatomical variants, and compare tract organizations across subjects in a way that is robust to individual variability. This approach enables both quantitative assessment of tract properties and visualization of complex fiber patterns, providing new insights into white matter organization and potential biomarkers for neurological conditions.

We applied and evaluated this framework to analyze thalamic white matter pathways, as thalamic structural alterations following TBI have been associated with executive dysfunction, cognitive impairment, and poor functional outcomes, including PTE \cite{little2010}. Studies have shown that selective thalamic neuronal loss occurs in chronic TBI and may play a critical role in recovery trajectories and epileptogenic network formation \cite{grossman2012}. The thalamus, through its extensive cortical connections, serves as a crucial hub in neural networks, making it particularly relevant for understanding TBI and PTE pathology as well as potential recovery mechanisms. By analyzing individual streamline patterns and their organizational properties within each thalamic bundle, our framework may capture subtle alterations in white matter architecture that are not detectable through conventional tract-averaged approaches, potentially providing more sensitive markers of injury-related changes and network reorganization.

The goal of this study is to apply our analytical framework to thalamic white matter pathways to detect subtle microstructural and organizational changes following TBI. By leveraging individual streamline characteristics and intra-bundle connectivity patterns, we aim to identify structural alterations that may contribute to executive dysfunction, cognitive impairment, and seizure susceptibility in PTE. This approach may provide more sensitive biomarkers of injury-related changes and network reorganization, offering new insights into the role of the thalamus in TBI pathology and recovery.

\section{Method}

\subsection{Data Acquisition}

Multi-center 3T MRI data, including diffusion tensor imaging (DTI), was acquired as part of the Transforming Research and Clinical Knowledge in TBI (TRACK-TBI) dataset from 11 Level 1 trauma centers using a standardized protocol \cite{manley2017tracktbi}. Whole-brain DTI was obtained using a spin-echo echoplanar sequence with 64 isotropic diffusion directions (b = 1300 s/mm²) and 8 b = 0 s/mm² acquisitions, with an isotropic voxel resolution of 2.7 mm.

All procedures in the original TRACK-TBI study were conducted under IRB approval at participating institutions, as described in their protocol \cite{manley2017tracktbi}.

\subsection{Preprocessing}

Standard DWI preprocessing was performed using a combination of the FMRIB Software Library (FSL) Diffusion Toolbox ~\cite{jenkinson2012}, MRtrix ~\cite{tournier2019mrtrix}, and the Quantitative Imaging Toolkit (QIT) ~\cite{cabeen2018qit}. The DWI images were denoised using Multichannel Principal Component Analysis (MP-PCA) ~\cite{veraart2016mp_pca, cordero2019gibbs} and corrected for Gibbs ringing artifacts ~\cite{kellner2016gibbs}, eddy current artifacts ~\cite{andersson2016eddy}, and bias field distortions using the N4 algorithm ~\cite{tustison2019n4}. Skull stripping was performed with FSL BET ~\cite{smith2002bet}. Diffusion tensors were estimated using FSL DTIFIT, and multi-tensor models (ball-and-sticks) were applied using FSL BEDPOSTX ~\cite{behrens2007probabilistic}. 

\subsection{Tractography}

We performed quantitative tractography analysis to characterize 20 thalamic white matter bundles bilaterally, connecting the thalamus to various cortical regions (Table 1). Bundle-specific tractography was performed using an atlas-based approach with the IIT ICBM diffusion MRI template \cite{varentsova2014}. Bundle definitions were created using group-averaged multifiber models, with regions of interest (ROIs) delineated based on the anatomical template parcellation. Following diffeomorphic registration of subject FA maps to template space using ANTs, the bundle delineation masks were transformed to subject native space. A hybrid reinforcement tractography approach was then applied using the ball-and-sticks models with parameters including a maximum turning angle of 75°, minimum volume fraction of 0.05, 1mm step size, and minimum length of 10mm. The resulting streamlines for each bundle were saved in curves.vtk.gz format for subsequent analysis.

\begin{table}[h]
\caption{Tract Names and Their Descriptions}
    \centering
    \begin{tabular}{ll} 
        \textbf{Tract Name} & \textbf{Description} \\ 
        pcc         & posterior cingulate connections \\
        pole        & temporal pole connections \\
        stem        & brainstem connections \\
        dlpfc       & dorsolateral prefrontal cortex connections \\
        dmpfc       & dorsomedial prefrontal cortex connections \\
        occ         & occipital lobe connections \\
        parcen      & central parietal cortex connections \\
        parinf      & inferior parietal cortex connections \\
        parmar      & marginal parietal cortex connections \\
        parmed      & medial parietal cortex connections \\
        parsup      & superior parietal cortex connections \\
        pma         & premotor area connections \\
        postcentral & postcentral gyrus connections \\
        precentral  & precentral gyrus connections \\
        sma         & supplementary motor area connections \\
        tempinf     & inferior temporal connections \\
        tempmid     & middle temporal connections \\
        tempsup     & superior temporal cortex connections \\
        vlpfc       & ventrolateral prefrontal cortex connections \\
        vmpfc       & ventromedial prefrontal cortex connections \\
        \bottomrule
    \end{tabular}
    \label{tab:tracts}
\end{table}

\subsection{Fiber Property Analysis}
Individual streamlines within each thalamic bundle were characterized using two types of features:

Shape features: Fiber length computed as the sum of Euclidean distances between consecutive points. Mean curvature calculated from the rate of change of tangent vectors along the fiber. Dispersion measured as the mean distance of points from the fiber's center of mass.

Position features: Start point coordinates (x,y,z) indicating where the fiber begins. End point coordinates (x,y,z) indicating where the fiber terminates. Midpoint coordinates (x,y,z) representing the average position of all points. 

\subsection{Diffusion Maps for Dimensionality Reduction}
We employed diffusion maps to transform the 12-dimensional feature space (3 shape features and 9 position coordinates) into a lower-dimensional representation that preserves anatomically meaningful relationships between fibers. First, pairwise distances were computed between all fibers using their combined features. These distances were converted to similarities using a Gaussian kernel with automatically determined width (set to the median of distances), creating an n×n similarity matrix where n is the number of fibers. After normalizing this matrix, its eigendecomposition yielded embedding coordinates that preserve both local and global relationships between fibers. The final embedding reduced the dimensionality to 3 coordinates, chosen to capture the primary modes of variation while allowing visualization. The relative importance of these dimensions was assessed through analysis of the eigenvalue spectrum. Figure 1 provides an overview of the methodology used in this study, illustrated for one example subject.

\subsection{Statistics}
Statistical analysis was performed to assess correlations between tract features and Glasgow Outcome Scale – Extended (GOSE) scores, which were collected 6 months post-TBI to evaluate functional recovery. GOSE is an 8-point scale, where lower scores (1–4) indicate severe disability or death, while higher scores (5–8) reflect increasing levels of independence and good recovery \cite{ranson2019gose}. Pearson correlation coefficients were calculated between GOSE scores and various tract features, including diffusion map embeddings (mean, standard deviation, and range) and mean pairwise distances. Statistical significance was set at p\textless0.05. All statistical analyses were performed using Python with SciPy.

\section{Results}

\begin{figure*}[ht!]
    \centering
    \includegraphics[width=\textwidth]{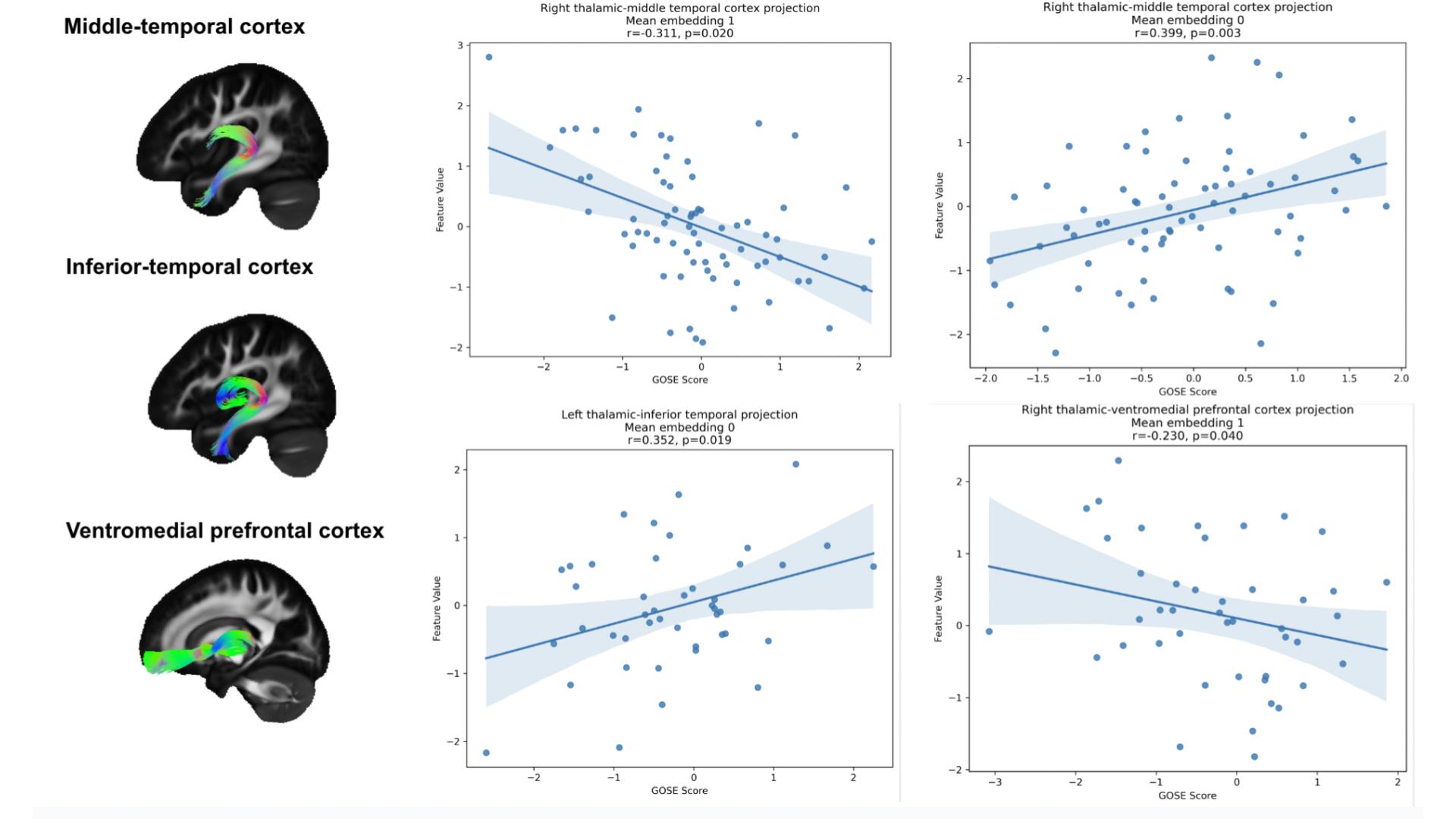}
    \centering
    \caption{Correlations between GOSE (Glasgow Outcome Scale - Extended) scores and diffusion-based embeddings of thalamocortical projections in different cortical regions. (Left) Streamline visualizations of thalamic projections to the middle-temporal cortex, inferior-temporal cortex, and ventromedial prefrontal cortex. (Right) Scatter plots showing relationships between GOSE scores and mean embeddings of thalamic projections to these cortical regions}
    \label{fig:example}
\end{figure*}

\subsection{Data Description}
The study included 75 subjects with mild TBI (mean age 40.5 ± 17.2 years, range 17-89 years). Glasgow Outcome Scale Extended (GOSE) scores had  a mean score of 6.26.  

\subsection{Feature Reduction Using Diffusion Maps}
Diffusion maps were applied to characterize the high-dimensional geometric features of each tract. For example, as seen in Figure 1 for the right thalamo-temporal mid tract (tempmid), the three-dimensional embedding revealed distinct patterns in fiber geometry. When colored by fiber length, the embedding showed a clear gradient from shorter fibers (60-70mm, shown in dark blue) to longer fibers (100-110mm, shown in yellow), suggesting that the first embedding dimension captured variations in fiber length. Similarly, when colored by curvature, the embedding displayed systematic variations in fiber shape, with curvature values ranging from 0.09 to 0.15. 

\subsection{Association of Thalamic Tract Properties with GOSE Scores}
Analysis of tract features revealed significant correlations between specific tract characteristics and GOSE scores. The strongest correlation was observed in the right thalamic-temporal mid pathway (tempmid), where the mean of the first embedding dimension showed a positive correlation with GOSE scores (r = 0.399, p = 0.003, n = 54). A similar correlation was found in the left thalamic-temporal inferior pathway (tempinf) (r = 0.352, p = 0.019) for the first embedding dimension. Additionally, the second embedding dimension of the right thalamic-temporal mid pathway (r = -0.311, p = 0.022) and the first embedding dimension of the left thalamic-temporal pole pathway (temppole) (r = -0.295, p = 0.034) were also significantly correlated with GOSE scores. Furthermore, the first embedding dimension of the right thalamic-ventromedial prefrontal cortex pathway (vmpfc) exhibited a negative correlation (r = -0.239, p = 0.049) with GOSE scores (Figure 2).

\section{DISCUSSION}

The results suggest that specific aspects of thalamic connectivity are linked to functional outcomes after TBI. Notably, the relationships observed in thalamic-temporal pathways (right temporal-mid and left temporal-inferior) highlight the potential role of thalamo-temporal connections in recovery. The positive correlations between embedding dimensions and GOSE scores in these tracts suggest that certain structural configurations may be associated with better functional outcomes.

The application of this method to thalamic white matter pathways is particularly relevant, given the role of the thalamus in cognitive function and network reorganization following TBI \cite{johnson2017}. Prior research has demonstrated selective thalamic neuronal loss in chronic TBI, suggesting that thalamic degeneration may contribute to persistent deficits and epileptogenesis \cite{garner2019imaging}. By analyzing individual streamline patterns and their relationships within each thalamic bundle, our framework may detect subtle alterations in white matter architecture that are not captured by traditional voxel-wise or tract-averaged diffusion metrics. Our findings suggest that diffusion map embeddings derived from streamline geometry capture meaningful structural variations, with certain features showing associations with functional outcomes as measured by GOSE scores.

A key advantage of our method is its ability to extract high-dimensional relationships between fibers while maintaining interpretability. Conventional DTI metrics such as FA and MD provide valuable information on microstructural integrity but can be challenging to interpret in the context of TBI due to their sensitivity to multiple pathological processes, including axonal damage, demyelination, inflammation, and edema. In contrast, our approach focuses on fiber geometry and spatial organization, which may provide complementary information on network-level changes following TBI. By applying diffusion maps, we reduce the dimensionality of complex fiber properties while retaining biologically relevant information, allowing for improved detection of structural variations associated with injury and recovery.

This study was limited to thalamic white matter pathways, and the generalizability of our findings to other brain regions remains to be explored. While the thalamus is a key hub in brain networks, TBI-induced white matter alterations are widespread and extend beyond thalamocortical connections \cite{little2010}. Future studies should apply this framework to additional white matter tracts implicated in TBI, including the corpus callosum and limbic pathways, to assess broader patterns of network disruption. For clinical translation, biomarkers derived from diffusion map embeddings must be comparable across subjects and populations. While our method employs atlas-based tractography and template registration to standardize bundle delineation, the diffusion map embedding process remains inherently subject-specific, meaning that embeddings computed in native space may not be directly comparable across individuals. This limitation hinders the establishment of normative values or clinical thresholds for disease severity and prognosis. Our future work will focus on developing population-level diffusion map templates that align embeddings to a standardized reference space, facilitating cross-subject comparisons and clinical integration. Additionally, longitudinal studies will investigate the stability of embeddings over time to determine their potential as biomarkers for disease progression or treatment response.

\section{CONCLUSIONS}

This study introduces a novel tractography analysis framework that leverages diffusion maps to extract meaningful white matter features at the individual streamline level. By applying this method to thalamic white matter pathways, we demonstrate its potential for detecting subtle structural changes following TBI. While our findings suggest that streamline-based features are associated with functional outcomes, further validation in larger, longitudinal cohorts is necessary. Future work should expand this approach to additional brain regions, integrate multi-modal imaging techniques, and explore its application in predictive modeling for TBI recovery and PTE risk.

\addtolength{\textheight}{-12cm}   % This command serves to balance the column lengths
                                  % on the last page of the document manually. It shortens
                                  % the textheight of the last page by a suitable amount.
                                  % This command does not take effect until the next page
                                  % so it should come on the page before the last. Make
                                  % sure that you do not shorten the textheight too much.

%%%%%%%%%%%%%%%%%%%%%%%%%%%%%%%%%%%%%%%%%%%%%%%%%%%%%%%%%%%%%%%%%%%%%%%%%%%%%%%%

%%%%%%%%%%%%%%%%%%%%%%%%%%%%%%%%%%%%%%%%%%%%%%%%%%%%%%%%%%%%%%%%%%%%%%%%%%%%%%%%

%%%%%%%%%%%%%%%%%%%%%%%%%%%%%%%%%%%%%%%%%%%%%%%%%%%%%%%%%%%%%%%%%%%%%%%%%%%%%%%%
\section*{APPENDIX}

\section*{ACKNOWLEDGMENT}

 Data used in the preparation of this article reside in the Department of Defense (DOD) and National Institutes of Health (NIH)-supported Federal Interagency Traumatic Brain Injury Research Informatics Systems (FITBIR) in FITBIR-STUDY 0000314 and FITBIR-STUDY 0000246. This manuscript reflects the authors' views and does not reflect the opinions or views of the DoD or the NIH.

%%%%%%%%%%%%%%%%%%%%%%%%%%%%%%%%%%%%%%%%%%%%%%%%%%%%%%%%%%%%%%%%%%%%%%%%%%%%%%%%

\end{document}